\def\year{2022}\relax
\documentclass[letterpaper]{article} 
\usepackage{aaai23}  
\usepackage{times}  
\usepackage{helvet}  
\usepackage{courier}  
\usepackage[hyphens]{url}  
\usepackage{graphicx} 
\usepackage{natbib}  
\usepackage{caption} 
\DeclareCaptionStyle{ruled}{labelfont=normalfont,labelsep=colon,strut=off} 
\usepackage{algorithm}
\usepackage{algorithmic}

\usepackage{booktabs}
\usepackage{comment}
\usepackage{color}
\usepackage{xcolor}
\usepackage{amsmath, array}
\usepackage{xspace}
\usepackage{placeins}
\usepackage{changepage}
\usepackage{longtable}
\usepackage{multirow}
\usepackage{siunitx}
\usepackage{supertabular}
\usepackage{enumitem}
\usepackage{textcomp} 
\usepackage{url} 
\usepackage{lipsum}

\usepackage{subfigure}
\usepackage{pifont}

\usepackage{booktabs}
\usepackage{tikz}
\usepackage{adjustbox}
\usepackage{colortbl}
\usepackage{multirow}
\usepackage{float} 

\definecolor{con}{HTML}{0070C0}
\definecolor{lab}{HTML}{FF0000}
\definecolor{tvt}{HTML}{FF7D18}
\definecolor{snp}{HTML}{FFFF00}
\definecolor{b60}{HTML}{7030A0}
\definecolor{ase}{HTML}{38C177}
\definecolor{lch}{HTML}{AB7942}

\usepackage{newfloat}
\usepackage{listings}
\floatstyle{ruled}
\newfloat{listing}{tb}{lst}{}
\floatname{listing}{Listing}
\title{\#TeamFollowBack: Detection \& Analysis of Follow Back Accounts on Social Media}
\author {
    Tuğrulcan Elmas \textsuperscript{\rm 1},
    Mathis Randl \textsuperscript{\rm 2},
    Youssef Attia \textsuperscript{\rm 2}
}
\affiliations {
    \textsuperscript{\rm 1} Indiana University Bloomington\\
    \textsuperscript{\rm 2} EPFL\\
    telmas@iu.edu, mathis.randl@epfl.ch, youssef.attia@epfl.ch
}

\definecolor{us}{RGB}{0, 128, 255} 
\definecolor{uk-fr}{RGB}{255, 0, 0} 
\definecolor{gr}{RGB}{0, 204, 0} 
\definecolor{id}{RGB}{255, 102, 0} 
\definecolor{ir}{RGB}{51, 153, 51} 
\definecolor{jp}{RGB}{204, 0, 102} 
\definecolor{kr}{RGB}{255, 255, 0} 
\definecolor{es}{RGB}{255, 153, 0} 
\definecolor{pk}{RGB}{0, 102, 204} 
\definecolor{ru}{RGB}{0, 51, 102} 
\definecolor{th}{RGB}{204, 204, 204} 
\definecolor{tr}{RGB}{255, 0, 0} 

\begin{document}

\newcommand{\Secref}[1]{Section~\ref{#1}}
\newcommand{\Figref}[1]{Figure~\ref{#1}}
\newcommand{\minisection}[1]{\vspace{0pt}\noindent\textbf{#1}}

\newcommand{\answerYes}[1]{\textcolor{blue}{#1}} 
\newcommand{\answerNo}[1]{\textcolor{teal}{#1}} 
\newcommand{\answerNA}[1]{\textcolor{gray}{#1}} 

\maketitle
\begin{abstract}
Follow back accounts inflate their follower counts by engaging in reciprocal followings. Such accounts manipulate the public and the algorithms by appearing more popular than they really are. Despite their potential harm, no studies have analyzed such accounts at scale. In this study, we present the first large-scale analysis of follow back accounts. We formally define follow back accounts and employ a honeypot approach to collect a dataset of such accounts on X (formerly Twitter). We discover and describe 12 communities of follow back accounts from 12 different countries, some of which exhibit clear political agenda. We analyze the characteristics of follow back accounts and report that they are newer, more engaging, and have more followings and followers. Finally, we propose a classifier for such accounts and report that models employing profile metadata and the ego network demonstrate promising results, although achieving high recall is challenging. Our study enhances understanding of the follow back accounts and discovering such accounts in the wild.
\end{abstract}
\section{Introduction}
\label{sec:intro}

Social media algorithms and the public often favor popular social media users due to their perceived influence. Thus, many users employ growth strategies to increase their number of subscribers or followers to appear more popular than they might actually be. One such growth strategy is to follow back anyone who follows them. Such accounts that promise reciprocal follows are named follow back accounts. Such accounts often use common hashtags like \#FollowBack, \#TeamFollowBack, \#Follow4Follow, or \#F4F in their tweets or profile descriptions to find and follow each other and create communities. Some may also use bots to automate follow backs.

Follow back accounts jeopardize the integrity of the social media platforms as they inflate the follower counts of themselves and other accounts. When an account accumulates followers primarily through follow backs, it suggests that it is not popular due to the merit of its content. This can mislead those unaware of the account's growth strategy into overestimating its true influence or expertise.

Follow back accounts may have the potential to cause harm if they also have malicious goals. If an account with an inflated follower count shares misinformation, the perceived credibility from the high follower number can lead more people to take the information at face value. Additionally, follow back accounts may sometimes follow spam or fake profiles which may inadvertently amplify such harmful profiles. Even though such accounts have initially benign goals despite their misleading follower counts, they may be later sold to adversaries who exploit their large audience to spread misleading content~\cite{elmas2020misleading}.

Despite their potential harm, such accounts are understudied. We fill this gap and present the first large-scale study of follow back accounts. We specifically focus the accounts on X, formerly Twitter, due to the data availability. Our contributions are formally defining the follow back behavior, establishing a ground-truth dataset using a honeypot approach, characterizing the follow back accounts and their communities, and proposing a classifier. We conducted our analysis in a global scope, without limiting it to a specific country.

\subsubsection{Summary of Findings:} We discovered 2759 follow back accounts and 12 follow back communities from 12 different countries. Some of these communities explicitly promote governments or political parties, suggesting that they may be used to influence public opinion. We found that follow back accounts are newer, have more followers and followings, and create and receive more engagements. Communities with higher follow back ratios also tend to observe those behaviors. Some communities abuse the platform by employing coordination, automation, and follow trains. However, it appears that X does not strongly enforce its policy on the accounts as only 6.3\% of the follow back accounts were suspended during the 2 years we collected the data. 

We found that a tabular data classifier using features created from profile metadata, the last tweets, and the ego network of the users can classify follow back accounts with moderate success. However, it does not generalize to all communities and reports low recall. Our approach may aid researchers in rooting out follow back accounts in the wild, and help them study coordination further.
\section{Related Work}
\label{sec:related}

\minisection{Social Media Manipulation:} There is extensive research on manipulation strategies and the actors involved. For instance, adversaries including government entities employ political troll accounts to influence public opinion and sway elections~\cite{zannettou2019let,balasubramanian2022leaders}. Automated accounts, and bots, amplify social media posts by reposting them~\cite{chavoshi2016identifying, elmas2022characterizing}. Coordinated users spread conspiracy theories and misinform the public~\cite{wang2023identifying}.

Illicitly growing social media accounts is also a manipulation strategy. Adversaries may grow their accounts by buying fake followers~\cite{fameforsale}, buying and repurposing entire accounts~\cite{elmas2020misleading}, riding "follow trains" (i.e., being recommended by other accounts explicitly)~\cite{torres2020manufacture}, maintaining backup accounts~\cite{merhi2023information}, or authoring viral tweets~\cite{elmas2023measuring}. Some social media accounts abuse reciprocity to illicitly grow their accounts. For instance, users sign up for illicit schemes so that their account will be used to promote others in exchange for the other accounts in the scheme will promote them by means of likes and reposts~\cite{weerasinghe2020pod}. 

Follow back accounts are such accounts that promise and do follow whoever follows them. The presence of such accounts is reported in the previous work. For instance, Wang et al.~\citeyear{wang2023identifying} observed that there were such accounts among QAnon supporters but the QAnon clusters they analyzed were not predominantly composed of them. Beers et al.~\citeyear{beers2023followback} propose "coengagement transformation" to visualize coengagement networks and argue that it can effectively visualize communities of accounts who follow each other. They reported that such accounts are used to show support for the presidential candidates Trump and Biden, although they did not extensively analyze those accounts. To the best of our knowledge, our study is the first to systematically analyze follow back accounts. We do this by the honeypot method, which was shown to be reliable in collecting the data of spammers~\cite{lee2011seven}.

The closest work to ours is by Mosleh et al.~\cite{MoslehPartisanShip} who analyzed follow back behavior in the context of hyper-partisan environments. They create honeypot accounts that identify themselves as Democrat or Republican and follow other users. They found that the users who are aligned with the same party of the bots following them are three times more likely to follow back. Although our approach in data collection is the same (i.e., we also created honeypot accounts to send follows to users), our work distinguishes itself by focusing on follow back accounts from a broader perspective instead of limiting it to hyper-partisan users within the context of U.S. politics and going beyond by analyzing and proposing a method to classify such accounts. 

\minisection{Follow-back Prediction:} Predicting follow-back accounts is similar to link prediction and reciprocity prediction which studies the underlying factors behind one or two-way link formation. Link prediction is predicting if two nodes will be linked or if the link between two nodes is missing~\cite{lu2011link}. It is studied extensively for social networks~\cite{xu2019link, kuo2013unsupervised, gao2015link,el2022twhin} including follower networks on Twitter~\cite{martinvcic2017link,valverde2013exploiting,quercia2012tweetlda,yuan2014exploiting}. Reciprocity prediction is different than link prediction as it predicts two-way link formations given one-way links~\cite{lou2013learning}. It is studied in the context of citation network~\cite{daud2017will} and social networks on Twitter~\cite{cheng2011predicting}. Hopcroft et al.~\citeyear{fbpred} studied the factors in reciprocal following and found that homophily is an important factor in reciprocal followings. Our study is different as we focus on detecting accounts that indiscriminately follow anyone to gain follow-backs in exchange instead of reciprocity among regular users.
\section{Definition}

We define the follow back behavior as \textit{engaging in reciprocal followings in order to mutually inflate follower counts}. There may be multiple types of follow back behavior: accounts may be \textit{actively} growing their account by sending follow requests hoping to get follow backs in return. They may also \textit{passively} engage in such behavior by following anyone who follow them first, even though they do not actively send follows. We define a follow back account as an account that observes follow back behavior with high probability. That is, the account will indiscriminately reciprocate by following back those who follow them, unless under exceptional circumstances such as being inactive on the platform or denying follow backs to malicious accounts or accounts in the opposite end of the political spectrum.

Accounts engaging in follow backs may be active, passive, or might have discontinued this behavior after achieving a significant audience size. Ideally, we can identify follow back accounts and whether they are active, passive, or discontinued in this regard by analyzing the frequency and timing of their follow requests. However, to the best of our knowledge, no social media platform provides such data. Thus, our study focuses on accounts that are at least passively engaging in follow backs, i.e., they follow back anyone who follows them but may not be proactively seeking to expand their audience. We now describe our ground truth data collection methodology tailored to this scope.  

\section{Data Collection}

\subsection{Ground Truth}

We assume that a follow back account that passively engages in follow back behavior is likely to follow back any account that follows them. By this assumption, we created five honeypot accounts that are blank, uninteresting profiles. That is, the accounts have empty profile attributes (description, location, home page, etc.). They do not have a profile photo or a background image. They do not have any tweets. Their name and screen names are random strings. We follow other accounts and log the time of the followings and the time of the follow back if we receive one. We do this by collecting the followings and followers of the honeypot accounts every 5 minutes. We assume the accounts which followed back or unsolicitedly followed the honeypots are follow back accounts with high probability. Our approach ensures our accounts lack political bias, malice, or fake identity since they do not assume any identity, making us able to attract follow back accounts that avoid spam or differing political stances.

Our ground truth data collection is not free of limitations. Firstly, it may suffer from recall: follow back accounts may not necessarily follow our accounts back. We acknowledge this limitation and argue that our main focus is to analyze the characteristics of the follow back accounts and not provide social media platforms with a detection methodology with a high recall. Thus, the number of follow back accounts we capture with this approach may suffice for our objective. 

\minisection{Do Not Follow Back Accounts:} Although our honeypot method is more reliable than hand labeling follow back accounts, it is still not perfect. First, the accounts that follow our accounts back may not be consistent in their behavior: they may not be following anyone who follow back even though they followed ours. Secondly, we also aim to follow back accounts that may be malicious or automated. However, the follow back accounts collected by our initial approach are not necessarily automated; they may be following anyone who follows them by hand. To capture accounts whose follow back are not coincidental and also to provide evidence that some of those accounts are malicious bot accounts, we introduce a new methodology. We created five additional honeypot accounts named "DO NOT FOLLOW BACK". The accounts have a profile picture of a stop sign. Their description field reads as ``We are EPFL researchers. You recently followed one of our bots. Please do NOT follow this account if you are NOT a malicious bot." We created two such accounts in English, and one in Japanese, Iranian, and Indonesian as the users tweeting in those languages had a substantial size in our dataset. We send follows to the follow back accounts we previously found, using the honeypot account created in their tweeting language if available. We assume that the accounts who followed those accounts may be malicious and may be automating their followings with high probability as they follow us despite our warning. Note that we assume the accounts read the warning or at least see the stop sign, as the X's Web Interface or the mobile app shows both the name and the description field of the user who sent the follow.

\subsection{Sampling Methodology}

Follow-back is a rare behavior~\cite{gtChen}. Sending follow requests is limited by Twitter T.O.S which allows only 400 requests per day. It is also labor-intensive. This makes finding follow-back accounts in the wild a challenge. Therefore, we employ two samples: a random sample where follow-back accounts are rare and a snowball sample which has more positives but may be biased towards certain communities of accounts, which we mitigate in our analysis. 


\minisection{Random Sample}: We downloaded the 1\% random sample of all tweets posted in the first half of 2020, which is the time we started this study, from the Internet Archive~\cite{team2020archive} and compiled the list of users who posted at least once. Since the users should be sufficiently active (expected to have at least 100 tweets) to appear in this dataset, the data is biased towards the users that are somewhat active in the platform. We followed a random set of 4246 users among them and received 142 follow backs (3.34\%). 

\minisection{Snowball Sample}: As the number of follow back accounts discovered in the random sample is too few for detection and analysis, we use snowball sampling to increase the number of positives. We send follow requests to a random sample of followers of the 15 randomly selected follow back accounts discovered in the random sample. We could not do this for all 142 accounts due to time constraints. We sent 3577 follower requests and got 1294 follows back (36\%). We also tested an approach where we send follow requests to accounts that have a follower-to-friend ratio between 0.95 and 1.0. This is because previous work states that follow back accounts have a nearly 1:1 followers-to-following ratio~\cite{torres2020manufacture,beers2023followback}. Interestingly, this did not substantially increase the rate of follow backs: We sent 2969 follow requests and got 1185 follow backs, which equals to a follow back rate of 39.9\%. The snowball sample helped us to discover communities of follow back accounts. Our caveat is that the sample is also biased towards certain communities. To mitigate this bias, we report results separately for each user group. We also acknowledge that the sample is not exhaustive and may not contain all the follow back communities. We have 6546 follows and received 2564 follow-backs (39.17\%) in the snowball sample in total. 

Our initial ground truth dataset consists of 8654 negatives and 2759 follow backs (23.36\%), totaling 11413 accounts. We followed 2628 of them for the second time from our ``Do Not Follow Back" accounts and received 1156 follow backs in total (44\%). This may indicate that nearly half of the follow back accounts may be automating the follow backs. 131 accounts became unreachable (either suspended or went private) later so we could not follow them in this second step.

\minisection{Additional Data:} To analyze the accounts and to build a classification methodology, we collected all of their profile data, their last 200 tweets, their friends, and their followers.

\section{Communities of Follow Back Accounts}
\label{sec:communities}


Multiple follow back accounts may form follow back communities. Those communities often have mutual objectives and interests that would help us to identify why the accounts observe follow back behavior. However, not all communities of accounts reciprocally follow each other are follow back communities. To identify a community as a follow back community, we require it to meet certain criteria. Firstly, it must have a substantial size. Additionally, it should maintain a minimum percentage of follow back accounts, referred to as the 'follow back ratio,' defined by a specific threshold. We believe that the follow back ratio is indicative of the likelihood that a community is indeed a follow back community and that the accounts within it are follow back accounts. We now explain our methodology for identifying follow back communities and describe the communities we found.

We build the social network of the accounts in our dataset using their follower and following data. We then detected the communities using the Louvain method~\cite{blondel2008fast}. We found 12 communities with a substantial size, with the smallest having 133 accounts. We merged the rest of the accounts and considered them to have no community, named None or the control group. The modularity is 0.739 which indicates a strong community structure. The lowest follow back ratio is 10\%. 

\begin{figure}[hbt]
    \centering
    \includegraphics[width=0.85\columnwidth]{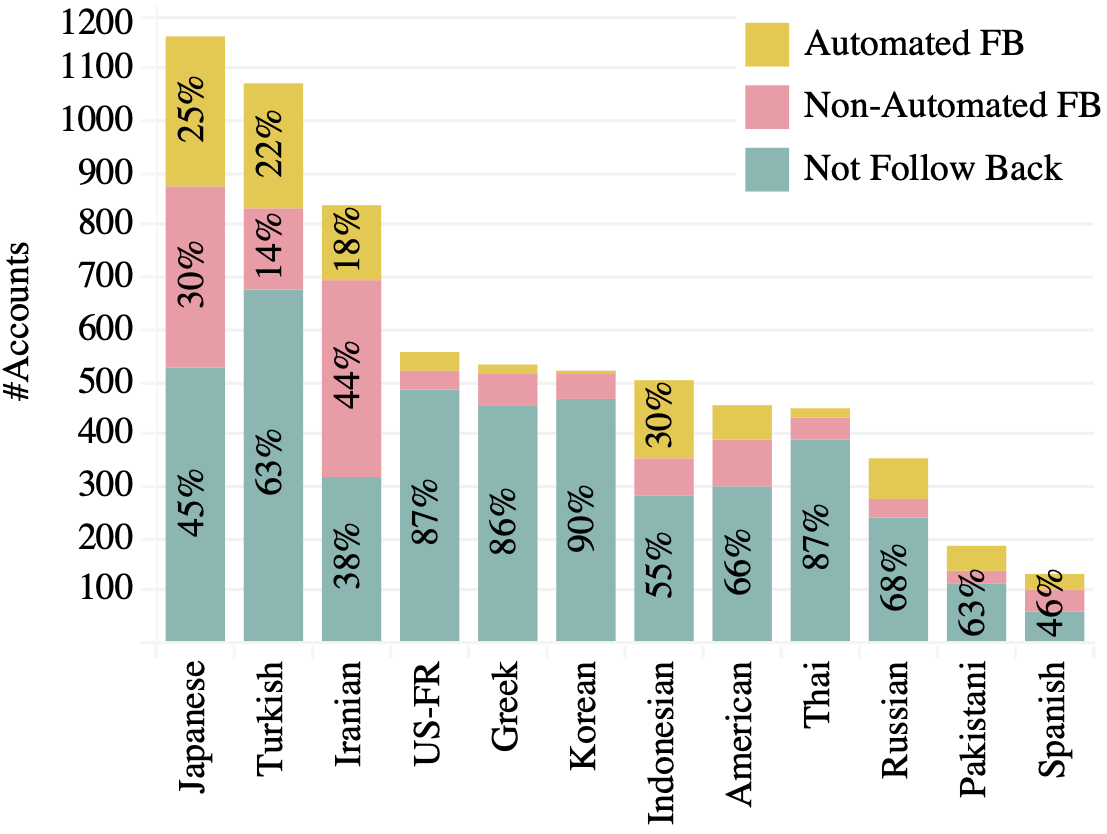}
    \caption{Communities, ordered by their size.} 
    \label{fig:communities}
\end{figure}

We name the communities based on the predominant country they appear or claim to be from. We estimate the country by manually inspecting the account names, descriptions, self-reported locations, and the languages used in each community. \Figref{fig:communities} shows the communities, the number of accounts and the percentage of likely automated or non-automated follow back accounts as well as the non follow back accounts. The follow back ratios vary between 62\% and 10\%. Japanese, Iranian, and Spanish are clearly follow back communities as the majority of the users followed back our honeypots. Turkish, Pakistani, American, and Russian communities have a substantial amount of follow back accounts too. Meanwhile US-FR, Greek, Korean, and Thai communities have a very low follow back rate, 14\% at most.

All communities' follow graphs have a reciprocity rate (i.e., the ratio of the number of edges pointing in both directions to the total number of edges) of at least 97\% with the exception of Thai community which has 56\%. We inspected this community and found that although it has a central network with very high reciprocity, it also has a substantial number of peripheral nodes that have many incoming links from the central network but do not have any outcoming links. The accounts in the central network may be used as fake followers to promote the peripheral nodes. 


\subsection{Semantic Characteristics of Communities}

\begin{table*}[t]
    \small
    \centering
    \begin{tabular}{p{.07\textwidth}|p{.43\textwidth}|p{.42\textwidth}}
        \toprule
        &  \textbf{Hashtags} & \textbf{Retweeted Users} \\
        \midrule
        \tikz\draw[black, fill=us] (0,0) circle (.75ex);  \texttt{US} & \#DemVoice1 (115), \#FBR (93), \#StrongerTogether (83), \#FBRParty (79), \#ONEV1 (71) & @joncoopertweets (144), @donwinslow (131), \newline @mmpadellan (130), @OccupyDemocrats (125), \newline @MeidasTouch (115) \\ 
        \tikz\draw[black, fill=uk-fr] (0,0) circle (.75ex);  \texttt{US-FR} & \#dogs (95), \#dogsoftwitter (94), \#COVID19 (58), \#dog (56), \#Zemmour (55) & @ZemmourEric (37), @F\_Desouche (36), @f\_philippot (32), @JuniorGuibole (31), @DamienRieu (27) \\
        \tikz\draw[black, fill=gr] (0,0) circle (.75ex);  \texttt{GR} & \#vipICU (109), \#KINAL (93), \#Mitsotakis (90), \newline \#Zaralikos (90), \#antireport (84) & @PGPapanikolaou (134), @KostasVaxevanis (103), @To\_pouli\_tou\_Ro (103), @kyrosgranazis\_ (96), \newline @zaralikos (89) \\
        \tikz\draw[black, fill=id] (0,0) circle (.75ex);  \texttt{ID} & \#programfolback (128), \#FollowFirst (115), \#RedAndWhiteMovement (82), \#FriendsOfTheNKRI (81), \newline \#ContinueRTsetiaNKRI (78) & @FerdinandHaean3 (132), @\_\_AnakKolong (118), @Dennysiregar7 (116), @jokowi (111), @Nadjib\_178 (107) \\
        \tikz\draw[black, fill=ir] (0,0) circle (.75ex);  \texttt{IR} & \#RetweetPlease (95), \#Retweet (62), \#Rashto (55), \#Isfahan (48), \#FollowBack (48) & @ZeddeEjtemaei (34), @ghasedakesepid (32), @YASERkhobpor (29), @DJ\_PashmackZzz (29), \newline @amiralisimpson (28) \\
        \tikz\draw[black, fill=jp] (0,0) circle (.75ex);  \texttt{JP} & \#SpreadPlease (281), \#MutualFollow (268), \#FollowBack100 (240), \#FollowBack (222), \#MutualFollow100 (194) & @aachaniina (42), @Awakend\_Citizen (41), @Kimura\_takuya20 (30), @chame\_kigyou (30), @shun1730 (29) \\
        \tikz\draw[black, fill=kr] (0,0) circle (.75ex);  \texttt{KR} & \#BTS (287), \#JIMIN (240), \#BTSARMY (146), \#BTS\_Butter (146), \#SUGA (126) & @BTS\_twt (348), @bts\_bighit (213), @jhoprs (127), @Univers\_Bangtan (125), @taeteland (109) \\
        \tikz\draw[black, fill=es] (0,0) circle (.75ex);  \texttt{ES} & \#OnlyVoxRemains (41), \#FirstVox (31), \#TeamVox (29), \#FollowMeAndIFollowYou (23), \#VoxGirls52 (21) & @Macarena\_Olona (36), @Santi\_ABASCAL (34), @Alvisepf (30), @ivanedlm (27), @CristinaSegui\_ (26) \\
        \tikz\draw[black, fill=pk] (0,0) circle (.75ex);  \texttt{PK} & \#Pakistan (57), \#BBN (42), \#NewProfilePic (37), \#LongLivePakistan (34), \#Sialkot (31) & @ImranKhanPTI (40), @SHABAZGIL (33), @WailaHu (27), @SdqJaan (26), @p4pakipower (24) \\
        \tikz\draw[black, fill=ru] (0,0) circle (.75ex);  \texttt{RU} & \#Russia (19), \#MyTwitterAnniversary (14), \#photography (14), \#nature (13), \#NaturePhotography (13) & @spacelordrock (58), @Snaiper41 (57), @EFn6oSqE2WMu0CZ (53), @b\_bratstvoBron (51), @Lora020563 (50) \\
        \tikz\draw[black, fill=th] (0,0) circle (.75ex);  \texttt{TH} & \#WeLoveTheMonarchy (106), \#COVID19 (97), \#LongLiveTheKing (81), \#NewProfilePic (67), \newline \#FathersDay (61) & @jjookklong3 (69), @huang\_huixian (64), @vnomenon (61), @political\_drama (59), @H2O\_Whan (54) \\
        \tikz\draw[black, fill=tr] (0,0) circle (.75ex);  \texttt{TR} & \#IStandByMyState (352), \#GoodFriday (242), \#YouWillBeAccountable (236), \#NationalTechnologyMove (199), \#ContinueToSpoilTheGames (190) & @RTErdogan (702), @06melihgokcek (445), @suleymansoylu (399), @zekibahce (283), @SireneOznur (236) \\
        \bottomrule
    \end{tabular}
    \caption{The most frequently used hashtags (translated to English) and retweeted users in the communities \label{tab:communities}}
\end{table*}

We describe the communities in terms of their content by presenting their most popular hashtags and users. We determine the popularity of a hashtag or a user by computing the number of users mentioning or retweeting them. Table \ref{tab:communities} shows these entities and the number of accounts mentioning them. We observe that the communities have either a political aspect i.e., they are partisan echo chambers, such as the Turkish and the American communities, or they have a commercial aspect and mainly focus on promoting each other, such as the Japanese and the Iranian community. Some communities observe both characteristics (e.g., the Indonesian). We now describe each community in detail.

\minisection{Japanese}: A community of accounts that self-state they offer account promotion services, which may indicate that they have a strong commercial aspect. Their most popular hashtags are "Please Retweet", "Mutual Follow", "Follow100", and "Follow", showing that the community is clearly a follow back community. This is further corroborated by the fact that almost half of the follow back accounts in the community also followed our "Do Not Follow Back" account despite our warning that was written in Japanese. Some users used automation tools as well. 74 accounts used twittbot.net, 25 accounts used botbird.net, and 24 accounts used social-dog.net to automate at least one tweet. 

\minisection{Turkish:} Accounts that self-state to be aligned with the Turkish president Erdogan and the ruling party AKP. Their most popular hashtags have a pro-government stance. They use the hashtag \#NationalistAccountsAreTogether" (\#MilliHesaplarYanYana) to find and follow each other. They promote state officials like Erdogan and Suleyman Soylu (Minister of the Interior) as well as pro-AKP influencers such as Melih Gokcek, Zeki Bahce, and Oznur Sirene. 

\minisection{American}: Accounts that self-state to be aligned with Democrats. They use \#FBR (which stands for Follow Back Resistance) and \#FBRParty to find and follow each other. 79 accounts (17\%) mention "Blue Wave" in their bio or use the wave emoji. There is strong support for the Black Lives Matter movement as 60 users use the hashtag \#BLM in their bio. They also widely share news from rawstory.com (52 users) which is a progressive news website. 

\minisection{Greek}: A community of accounts from Greece with diverse backgrounds. They frequently mention KINAL (PASOK), a minor opposition party that won 8\% of the votes in the 2019 Greek elections. Some accounts used the hashtag \#vipICU to criticize Mitsotakis and the government's unequal policies in ICU usage, which may suggest that the community is affiliated with the opposition.

\minisection{Indonesian}: Accounts that appear to be mostly composed of spam accounts that either retweet others or post "trains", i.e., a tweet that contains \#programfolback (Follow Back Program), profile handles of other users, and a viral video. As substantial, if not the majority, of the accounts in this community, are followed by both our initial and our "Do Not Follow Back" accounts, the community may be an Indonesian bot net that is used to promote others. Its third most popular hashtag \#RedAndWhiteMovement (\#GerakanMERAHPUTIH, a reference to the Indonesian flag) is a political hashtag shared with patriotic sentiment, praising and expressing love for Indonesia. The user they promote the most, @FerdinandHaean3, is an Indonesian political leader. These suggest that the community also has a political component.

\minisection{US-FR}: A mix of accounts that are from the U.S., France, the U.K., and Canada. The community does not appear to have a single agenda, but they widely promote far-right French politicians Eric Zemmour and Florian Philippot, the far-right media website Fdesouche. They also widely share dogs using the hashtag \#dogs (in English). The American accounts in this community state that they are conservatives. The hashtag COVID-19 was also widely used. We inspected the tweets containing this hashtag and found that some French-speaking accounts protested against the compulsory vaccination policy in France.

\minisection{Spanish:} Accounts from Spain that use the hashtag \#FollowMeAndIFollowYou (\#SiguemeYTeSigo) to find each other. They are mostly promoting the Spanish far-right political party Vox with hashtags such as \#OnlyVox (\#SoloQuedaVox) and their politicians such as Olona Macarena, Santi Abascal, Iván Espinosa de los Monteros.

\minisection{Russia:} Accounts that mostly retweet Russian propaganda accounts. Their top four most popular users were either suspended or inactive by 2023, which may be due to Twitter's interventions in Russian disinformation campaigns. 

\minisection{Pakistani:} Pakistani accounts using \#BBN ("Brotherhood Networking" in Urdu) to find each other. They are pro-government, as they use \#LongLivePakistan (\#PakistanZindabad) and retweet Imran Khan, the ex-president.

\minisection{Thai}: Thai accounts that appear to praise the king, and the kingdom, as their most popular hashtags is \#WeLoveTheMonarchy. 54 accounts also use the hashtag "Support 112", which is the Thai law that prohibits defaming or insulting the king. They retweet pro-king political accounts the most.

\minisection{Korean}: Fans of the popular K-Pop band BTS. They promote the accounts and hashtags related to the band.

\minisection{Iranian:} A community of accounts that appear to be mostly dedicated to account promotion and growth as their most popular hashtags indicate that the user seeks retweets. We could not observe or identify any common topic such as politics among the hashtags or the accounts promoted.

\section{Characterization}

\newcommand{\colwid}{90pt}

\begin{table}[ht]
\centering
\small
\resizebox{0.47\textwidth}{!}{
\begin{tabular}{p{\colwid}p{18pt}p{17pt}p{19pt}|p{19pt}p{10pt}}
\toprule
               Measure & \multicolumn{3}{c}{Binary} & \multicolumn{2}{c}{Community} \\
               &     FB &  Other &   Diff & Corr &    p \\
\midrule
\midrule
     Response (Median) &   7.66 &      - &   - &     -0.52 & 0.08 \\
    Age (Median) &   1.83 &   3.58 &  -1.75  &  -0.39 & 0.20 \\
    Followers (Median) &  3.2K &    785 &  2.44K &    0.06 & 0.86 \\
    Following (Median) &  3.4K &  778 &  2.64K &    0.04 &  0.90 \\
Followers/Age (Median) & 842 & 152 & 690 &     0.3 & 0.35 \\
Following/Age (Median) & 904 & 156 & 748 &    0.27 & 0.39 \\

    Reciprocity (Mean) &   0.83 &   0.48 &   0.35 &     0.5 &  0.10 \\
     Statuses (Median) &  5.5K &  6.1K &   -614 &    -0.28 & 0.37 \\

        Likes (Median) & 10.4K &  7.6K &  2.7K &    -0.12 &  0.70 \\
Statuses/Age (Median) &  1.49K &  1.24K &  245.4 &    0.13 & 0.69 \\
    Likes/Age (Median) &  3K &  1.6K &  1.4K &     0.25 & 0.44 \\

      Retweets\% (Mean) &   0.37 &   0.42 &  -0.05 &    -0.53 & 0.08 \\
Engagements (Median) &    356 &    141 &    215 &    0.74 & 0.01 \\
\bottomrule
\multicolumn{5}{l}{} 
\end{tabular}
}
\caption{Summary of the quantitative differences between follow back (FB) accounts and others, and the Pearson correlation between the follow back ratio and each measure. All differences between follow back accounts and others are statistically significant (p $<$ 0.05) except for Statuses, as indicated by either the Welch’s t-test for mean values or the Wilcoxon rank-sum test for median values. The p values for these tests are omitted from the table for brevity. Correlation coefficient p-values are presented in the last column.}
\label{tab:rq3summary}
\end{table}

In this section, we describe the characteristics of the follow back accounts, and their communities using a comparative approach, contrasting their behaviors with other accounts or communities. To facilitate comparisons, we introduce a measure for each behavior, computing the mean value for each group and using the median for measures susceptible to outliers. We then compare the group-wise differences, with emphasis on highlighting characteristics when the differences are both substantial and statistically significant.

We first do a \textit{binary analysis} and compare follow back accounts and other accounts. We report the difference in the mean/median value of each of these two groups. To test the statistical significance of the difference, we use Welch's t-test, which is more suitable when we compare mean values, and Wilcoxon rank-sum test, which is more suitable when we compare median values. We observe that all comparisons pass both of the tests except the Statuses (the number of all tweets of an account) that passes neither.

We then conduct a \textit{community-wise analysis}. We hypothesize that the measures that distinguish the follow back accounts from other accounts should correlate with the follow back ratio of communities. In essence, communities with a higher proportion of follow-back accounts should exhibit behaviors more specific to follow-back accounts. Thus, for each measure, we present the correlation between the mean/median of the measure and the follow back ratio (the percentage of follow back accounts in the community). We also report if some communities create outliers, and cause us to underestimate correlation. Notably, this is usually the case with Korean and Thai communities which are the communities with the lowest number and ratio of follow back accounts. We also report if the results differ whether we include only the follow back accounts in the community. We use the Pearson correlation coefficient to report the correlation between the follow back ratio and each measure. 


\begin{figure*}[htb]
    \centering
    \includegraphics[width=0.95\textwidth]{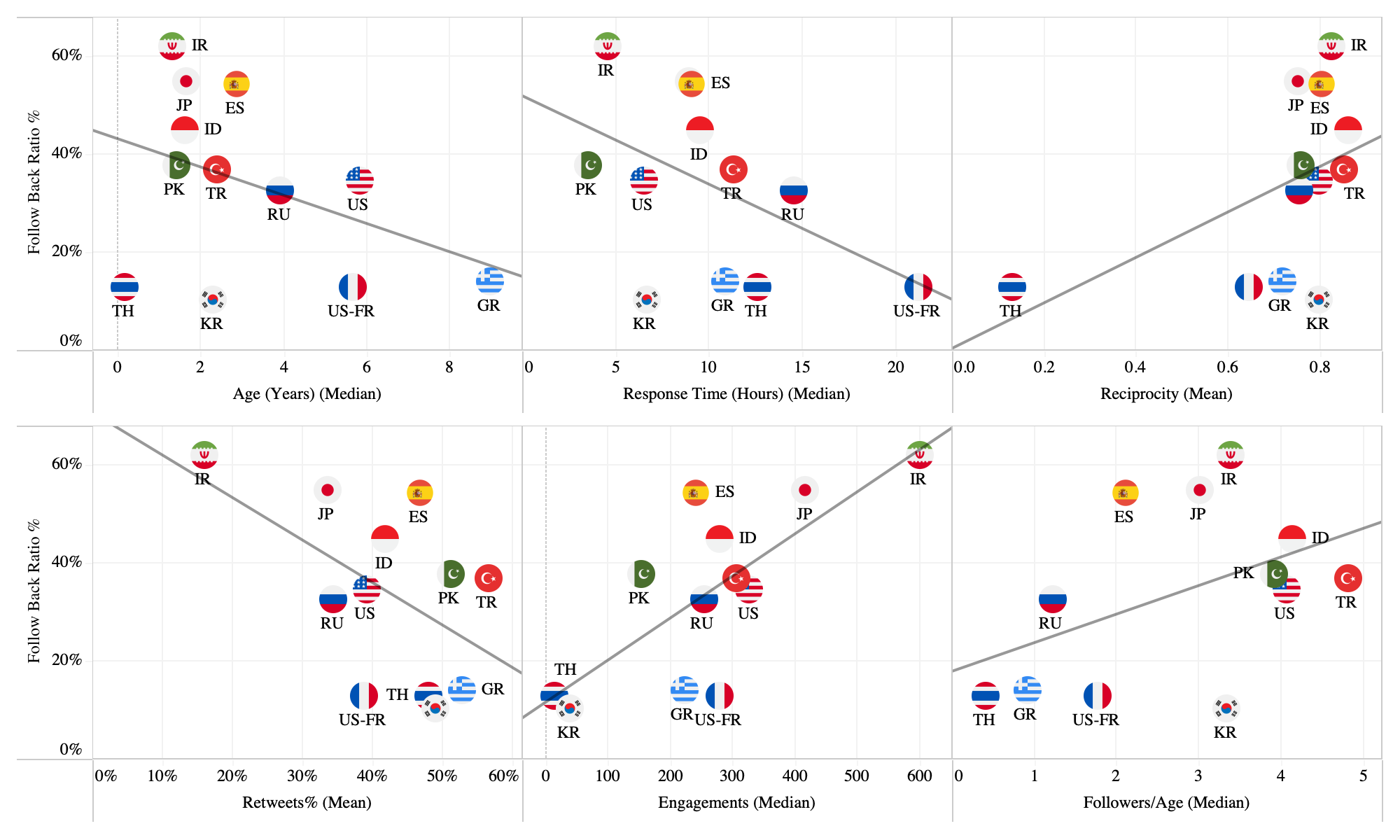}
    \caption{The characteristics vs. follow back ratios. The trend lines are computed using linear regression.} 
    \label{fig:community_characteristics}
\end{figure*}

\minisection{Age:} Follow back accounts are newer accounts and the communities they are prevalent in are younger to some extent. This may be because the newer accounts are more likely to be actively growing their community while the older accounts have already grown enough to engage in follow backs. It may be also because the older users consider their accounts to be valuable and are less likely to be involved in such schemes. To present this, we compute the account age using November 12th, 2021 as the pivot, which is the creation date of the newest account in the dataset. Its distribution is not normal and skewed towards the newer accounts. Thus, we report results using both the mean and the median.

The median/mean account age is 1.4/4.9 years for follow back accounts and 3.1/5.6 years for the others. The difference in median account age between follow back and other accounts in a community is more nuanced in communities where follow back account ratio is low, such as the Greek (8.6 vs 5.9 years), and the American (6.15 vs. 4.2 years) communities. This is in contrast to the communities with a high follow back ratio where the difference is lower, such as Japanese (1.82 vs. 1.56) and Iranian (1.45 vs. 1.28). Meanwhile, the follow back accounts in the American-French community are older (6.39 vs. 5.62).


The follow back ratio of a community is inversely correlated with the median account age in a community, including all or only the follow back accounts, as shown in \Figref{fig:community_characteristics} (Corr = -0.39, p = 0.2). However, the Thai and the Korean communities create outliers. The correlation between the followback ratio and the mean/median age becomes -0.79 in both cases ( p $<$ 0.01) when they are excluded. 

\minisection{Response Time:} Accounts in communities with a higher follow back ratio follow back our honeypots faster. To show this, we compute the response time as the duration between the time we followed the account and the time it followed us back. We use the median as the mean is susceptible to outliers (e.g., the max. response time is 229 days).

The median response time for all accounts is 7.6 hours. Meanwhile, only 22\% of the follow backs are within an hour, and only 10\% of the accounts responded in 5 minutes. This may suggest that follow backs are not automated with a simple approach in which the accounts follow back immediately. As there is no response time for non-follow back accounts, we proceed to compare the communities.

The median response times are inversely correlated with the percentage of follow back accounts in a community although the correlation is slightly low and not statistically significant (Corr = -0.52, p = 0.08). This is again due to Korean and Thai communities having lower response times despite having a low follow back ratio, and the correlation is higher and significant if they were excluded (Corr = -0.67, p = 0.03). The Pakistani community observes the lowest median response time (4h), followed by the Iranian (5h), American (5h) and Korean (6h) communities.

The Iranian and Japanese communities also stand out with the highest number of accounts having a response time within five minutes, including 120 (24\%) and 46 (7\%) such accounts respectively. There are also 54 accounts (2\%) that followed our accounts although we did not send them follow requests. They are discarded from this analysis as they do not have response time. 14 of them are from the Iranian, and 28 of them are from the Japanese community.


\minisection{Followings \& Followers:} The number of followings may indicate how aggressively the account aims to grow, while the number of followers may indicate the account's success in growing. As expected, follow back accounts observe those behavior, with inflated follower and following counts. The median followings for follow back accounts is 3415 (vs. 779), and the median number of followers is 3222 (vs. 785). We observe the same phenomena in each community. 

Most of the accounts (85\% of the follow back accounts and 91\% of all) do not pass beyond 10k follower counts, suggesting that follow back may not be effective in creating very popular accounts. However, the study may have a reverse survivorship bias: the accounts that became very popular may not be sending follow back anymore and may have already isolated themselves from their respective follow back communities, and did not show up in our dataset. 

We observe that there is no correlation between the mean/median follower or following counts with the follow back ratio of communities, even after filtering certain accounts and communities. This is because those values are rather influenced by the account age, i.e., older accounts have more followers and followings. The correlation between the account age and the followers/followings is 0.42 (p = 0.17 in both cases). We normalize these values by age and found that the follow back ratio is correlated with median following/age (Corr = 0.27, p = 0.39) and with the median followers/age (Corr = 0.3, p = 0.35) although the correlation is low and not significant. Excluding Korean or Thai communities or non-follow back accounts further decreases the correlation. We conclude that there is no meaningful relationship between the communities' following and follower counts and their follow back ratio. In other words, other factors than follow back ratio influence the follower and the following count of a community.

\minisection{Reciprocity:} As previously mentioned, the reciprocity of a directed graph is the ratio of bidirectional (mutual) links to the total number of links. This value is high for all communities (except for the Thai community) when we only consider the accounts in our dataset. This value is consistently high for all communities (except the Thai community) when considering only the accounts in our dataset. However, results vary slightly when we include accounts outside the dataset that have connections to the accounts within our dataset.

In this analysis, we shift our focus to the reciprocity of individual users, encompassing all their connections, even those not included in our ground truth dataset. We define the reciprocity of users as the Jaccard coefficients of their followings and followers, i.e., $\frac{|Followings \cap Followers|}{|Followings \cup Followers|}$. As anticipated, the mean reciprocity is higher among the follow back accounts overall (0.83 vs. 0.48), and in each community. The mean reciprocity in a community is correlated with the percentage of follow backs including all the accounts (Corr = 0.5, p = 0.1). The correlation increases to 0.66 (p = 0.036) when the Thai and the Korean communities are excluded. Interestingly, the correlation decreases dramatically when we only consider the follow back accounts in a community (Corr = 0.39, p = 0.21). This may suggest that follow back accounts across different communities have more or less the same level of reciprocity, and it is the other accounts that contribute to differences across communities.

\minisection{Activity:} Users may be actively using the platform, posting new content, and engaging with others by retweeting and liking. We measure the activity level of an account using its statuses count (which is the number of all tweets and retweets it posted) and likes count. We have mixed results: the median number of likes is higher on follow back accounts (10.4K vs. 7.6K) but the median number of statuses is higher on non-follow back accounts (6.1K vs. 5.5K). The results also differ across communities. While the follow back accounts have a higher median number of statuses in most of the communities, they have less in Indonesian, American, Russian, and Spanish communities. The median number of statuses of a community with the follow back ratio is correlated only if the Thai and the Korean communities are excluded (Corr = -0.54, p = 0.10). The correlation is slightly higher if we only consider the follow back accounts in each community (Corr = -0.59, p = 0.07). Interestingly, there is no correlation between the median likes count and the follow back ratio.

Similar to the case of followers and followings, the account age is highly correlated with the median statuses count (Corr = 0.75, p = 0.004), and the median likes count to some extent (Corr = 0.35, p = 0.25). To isolate the effect of age, we normalize the status count and likes count by dividing them by the account age. In this case, both the mean and the median Statuses/Age and Likes/Age are higher on follow back accounts, indicating that they are more active overall. However, there is no correlation between the follow back ratio of a community and its mean/median Statuses/Age or Likes/Age, whether we exclude the Thai and the Korean communities or include only the follow back accounts. We conclude that there is no meaningful relationship between the communities' activity level and their follow back ratio.

\minisection{Retweet Ratio:} Retweeting is a simpler function than creating authentic tweets. Thus, a high retweet ratio may be indicative of automated behavior. We compute the retweet ratio by dividing the number of retweets by the number of all tweets the user posted, limited to the last 200 tweets. 

Surprisingly, we found that the mean retweet ratio is lower on follow back accounts overall (37\% vs 42\%). The communities have mixed results: the follow back accounts in Japanese, American, Russian, and Pakistani communities have higher retweet ratio to their counterparts. There is an inverse correlation between the mean retweet ratio and the follow back ratio of a community (Corr = -0.52, p = 0.08), although it is mainly due to Iranian community that has an anomalously low mean retweet ratio (16\%). If this community is excluded, the mean retweet ratio of follow back accounts increases to 42.5\%, making them indistinguishable from the other accounts that have a mean retweet ratio of 43\%. The correlation also decreases to -0.28 (p = 0.4) if this community is excluded. The low retweet ratio in this community is due to the high reply ratio. We inspected a sample of Iranian tweets and found that they mostly consist of short interactions such as people thanking or affirming each other. It is not clear why this community engages in such interactions more frequently than the other communities. We conclude that there is no clear difference between retweet ratios among follow back accounts and communities.

\minisection{Engagements Received:} Engagements, which are likes and retweets, further boost an account's popularity and visibility. Follow back accounts receive more engagements, which may either because they are more successful in attracting them, or because they engage in other kinds of reciprocity abuse to inflate them. To show this, we compute the number of engagements an account receives by summing up all the likes and retweets it received. As this is computed over the last 200 tweets, it is not influenced by the account age. 

In general, both the mean and median values of engagements received are higher for follow-back accounts. These results hold for all communities, except for American and Japanese communities, where the differences are marginal. The median account engagement per community correlates with the follow back ratio (corr = 0.74, p = 0.006). This suggests that communities exhibiting high reciprocity in follows also tend to have high reciprocity in engagements.

\section{Platform Abuse}

In this section, we describe the behaviors related to platform abuse that were reported in the past work, to show the potential harm of follow back accounts. We also evaluate how the platform mitigates the problem by focusing on account suspensions.

\begin{figure}[!htb]
    \centering
    \includegraphics[width=0.5\textwidth]{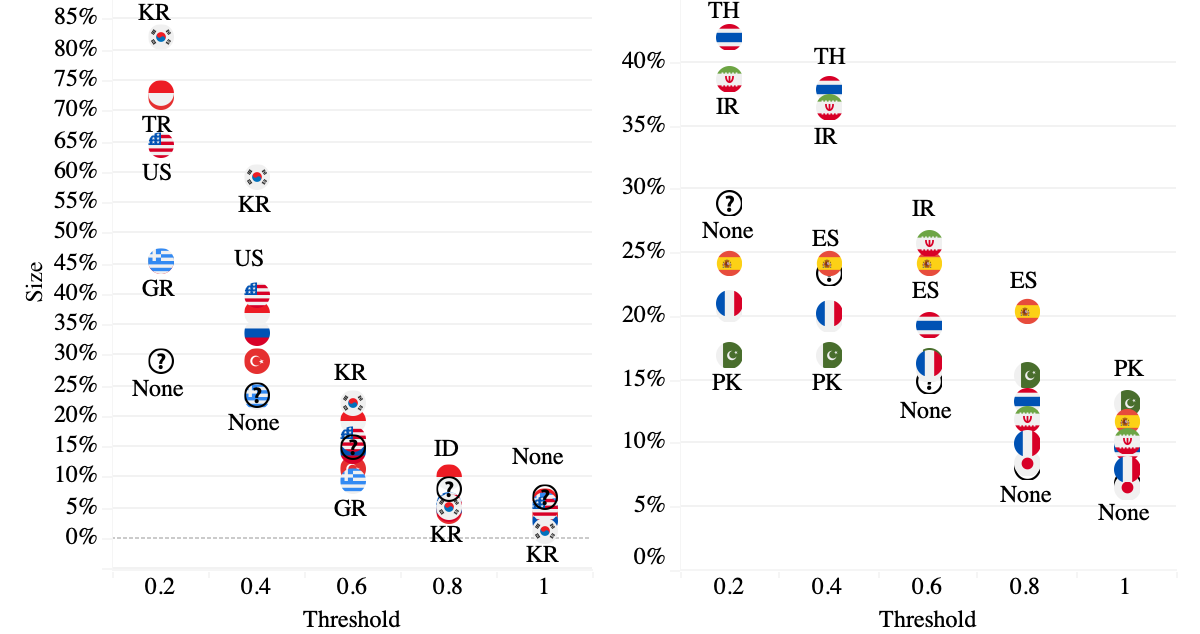}
    \caption{Level of coordination among communities: The communities depicted in the left plot exhibit a high number of weakly coordinated users but a low number of highly coordinated users. In contrast, the communities shown in the right plot have a relatively high number of highly correlated users but a low overall count of coordinated users.} 
    \label{fig:coordination}
\end{figure}

\minisection{Coordination:} We analyze the level of coordination within each community by the method proposed by Nizzoli et al.~\citeyear{nizzoli2021coordinated}. Our approach assumes that users promoting the same tweets, especially unpopular ones, are more likely to be coordinated. Thus, we treat each user as a TF-IDF weighted vector of the tweets its engaged with. Then, we compute cosine similarities between the user vectors and use it as a proxy for coordination between two users. We create the coordination network in which the nodes denote the users in a community, and edges denote the value of coordination (cosine similarity) between two users. Then, for each threshold of coordination value, we filter out the edges that have a lower value than the current threshold and compute the ratio of users that have at least one connection to the giant component of the network.

We observe two distinct patterns, and thus, present two plots containing the communities showing the respective pattern in \Figref{fig:coordination}. In the first pattern, the communities have a high number of weakly coordinated users but a very low number of highly coordinated users. Korean, Turkish, Japanese, American, and Greek communities observe this pattern. This may be because these communities have a common interest or agenda (e.g., retweeting the same politicians), but they promote a wide range of users. In the second pattern, the communities do not have a lot of coordinated users, but they have relatively high number of highly correlated users, making up between 7\% - 13\% of each community. Spanish, French, Pakistani, Iranian, and Thai communities observe this pattern. The two patterns are independent of the follow back ratio of the communities as both patterns feature communities with high and low follow back ratios. 

\minisection{Automation:} Assuming the accounts that followed back our "Do Not Follow Back Accounts" are automated, we compute the automation ratio among follow back accounts of each community. It is highest among the accounts in the Pakistani community (71\%), followed by the Russian (69\%), Indonesian (66.5\%), and Turkish (61\%) community. The communities with the lowest follow back ratio have very low automation ratios too, e.g., Korean has 2\% (only one account), Greek has 24\%, Thai has 28\%, and American-French has 46\%. Interestingly, the Iranian community, which has the highest follow back ratio, has only 28\% automation ratio. 

We repeat the same analysis as the previous section to see if the automated follow back accounts are different than the others. Overall, we do not observe clear differences. The median age of the automated accounts is roughly the same as the others (1.98 vs. 1.64). The median response time of automated follow back accounts is also roughly the same as the non-automated follow back accounts (8 hours). However, the response time from the automated follow back accounts contain fewer outliers and the distribution is more concentrated, i.e., the 75\% percentile is 29 hours for automated follow back accounts but 56 hours for non-automated accounts. The automated accounts have higher mean reciprocity (0.89 vs. 0.78), and higher median engagements received (431 vs. 354). A notable distinction is the retweet ratio, which is higher on average (40\% vs. 33\%). The median Followers/Age and Followings/Age are higher on automated follow back accounts (followings 1240 vs 650, followers 1190 vs 616). The results are statistically significant in all cases (p $<$ 0.05) except for the account age and the response time. We conclude that automated follow back accounts usually observe similar behavior to their non-automated counterparts. 

\minisection{Trains:} Follow trains are a social media practice where users promote (or get promoted by others in) long lists of profile handles, e.g., "Follow those accounts @1, @2,...". We analyze the train conducting (posting profile handles lists) and train riding (being mentioned in these lists) behavior of the communities. Similar to Torres et al.~\citeyear{torres2020manufacture}, we classify a tweet as a follow train if it is not a reply tweet and if it contains at least 5 profile handles. We selected this threshold as the tweets containing the "programfolback" (the most popular follow back hashtag) and the substring "promote" typically contains 5 handles, i.e., when we compute the number of handles in those tweets, we found that they most frequently promoted 5 handles. We compute the number of times a user conducts and rides a train.

We found that this behavior is extremely prevalent in the Pakistani community, as the mean number of rides per user is 13, and the mean number of conducts is 1.5. It is also widely observed in the Turkish (rides = 5.5, conducts = 0.8), Indonesian (rides = 5.5, conducts = 1), American (rides = 4, conducts = 0.5), and Spanish communities (rides = 2.2, conducts = 0.46). Interestingly, the American-French community has 0.5 train conducts, but very low rides per account 0.25, which may suggest that the community is mainly focused on promoting other accounts outside of the community. The follow trains are not observed in non-political communities such as Japanese and Iranian community despite their high follow back ratio. This suggests that follow trains are the most popular among partisan echo chambers.

\minisection{Suspensions:} Follow backs and other aggressive following strategies are explicitly prohibited on Twitter. However, the suspension rate is remarkably low. One year after the initial collection, in November 2022, only 401 accounts were suspended and 505 accounts were deleted. By July 2023, after Elon Musk purchased Twitter, 502 accounts were suspended and 722 accounts were deleted in total. 133 accounts were newly suspended and 32 accounts were reinstated. Of the 502 accounts that were suspended by 2023, 176 accounts followed our account back (6.7\% of all follow back accounts), 86 accounts also followed our "Do Not Follow Back" accounts (7.4\% of all automated follow back accounts), and 211 accounts were associated with one of the follow back communities even though they did not follow us back (4.9\%). The suspension rate in the random sample (excluding the accounts that followed us back) is 4.3\%. This shows that follow back accounts are only 36\% more likely to be suspended, suggesting that Twitter has not enforced its T.O.S. strongly on such accounts so far.

\section{Classification}

We now present our experimental results on classifying follow back accounts. Our objective is to understand which types of models are effective in detecting them. Furthermore, when combined with our data collection strategy, our classifiers can be employed to discover follow back communities for further study. We do not provide a framework for social media platforms to detect such accounts as they have access to more informative signals that we cannot obtain.

Our classification problem is to predict if a social media account is a follow back account or not. There are two approaches to this problem. First, we can model an account depending only on its data (e.g., profile features, its ego network, tweets). Second, we adopt a holistic approach by modeling an account within the context of the entire account network we collected. The latter is contingent on the amount and the quality of the data collected. Since it is unrealistic to collect the whole account network of a social media platform, we acknowledge this as a limitation.

Even when focused on a single account to create features, different types of data require different model considerations and are subject to different data collection limitations. The models with more flexible data requirements may be even preferable if their performance is not significantly lower than the best-performing models given the API constraints.

\minisection{Data:} Although our dataset contains 10,715 accounts for which we collected their tweet and network data, before they got suspended, only 6,881 nodes are directly connected to another node in our dataset, which we use in our experiments. We observe that 90\% of the accounts that are discarded are from the random set and almost all of them are non-follow back accounts. Thus, the remaining data is a more relevant subset for our analyses.

\minisection{Evaluation:} Follow back accounts are rare in the wild. Once an account is flagged as a follow back account, its classification can be verified by sending it a follow request. Additionally, the miss-classifications in aggregate may be less impactful in discovering follow back communities, i.e., it is less likely to miss-classify an entire group of accounts. Thus, we prefer models with high recall and sufficient precision.

We employ two different data-splitting strategies to create test sets. We first create the ``Random" set. We randomly sample 500 positives (which is \~20\% of all positives) and 500 negatives to maintain a balanced test set. However, this set places more emphasis on larger communities. To address this, we also create a ``Stratified" set wherein we select 25 positives and 25 negatives for each community, creating a 13 mini-test sets, each compromising 50 data points. We evaluate models in each of these mini-test sets and report the average score. This is to ensure the models generalize and do well equally across all communities.

\begin{table}[t]
    \centering
    \small
    \resizebox{0.48\textwidth}{!}{
    \begin{tabular}{p{90pt}S[table-format=1.2]S[table-format=1.2]S[table-format=1.2]S[table-format=1.2]S[table-format=1.2]S[table-format=1.2]}
       
        \toprule 
        \multicolumn{1}{c}{} & \multicolumn{3}{c}{\texttt{Stratified}} & \multicolumn{3}{c}{\texttt{Random}}\\
        \cmidrule(lr){2-4} \cmidrule(lr){5-7}
        Model  & {Prec.} & {Rec.} & {F1} & {Prec.} & {Rec.} & {F1}\\
        \midrule
        Profile &           0.55 &        0.16 &    0.23 &           0.70 &        0.24 &    0.36 \\
        Profile + Tweets &           0.58 &        0.15 &    0.22 &           0.77 &        0.23 &    0.35 \\
        \textbf{Profile + Tweets + Ego} &           0.79 &        \textbf{0.37} &    \textbf{0.47} &           0.82 &        0.53 &    0.65 \\

        \midrule 
        Handcrafted Network &           0.69 &        0.28 &    0.35 &           0.77 &        0.50 &    0.61 \\
        node2vec & 0.45 &        0.19 &    0.22 &           0.81 &        0.42 &    0.56 \\
        \midrule 
\textbf{P.T.E. + Network} &           \textbf{0.82} &        0.35 &    0.45 &           \textbf{0.84} &        0.58 &    \textbf{0.69} \\
        P.T.E. + node2vec & 0.64 &        0.32 &    0.39 &           \textbf{0.84} &        0.56 &    0.67 \\
        \midrule
        \textbf{GCN} & 0.59 & 0.21 & 0.29 & 0.68 & \textbf{0.70} & \textbf{0.69} \\
        \textbf{GAT} & 0.62 & 0.25 & 0.33 & 0.68 & 0.69 & \textbf{0.69} \\ 
        GraphSage & 0.60 & 0.30 & 0.38 & 0.75 & 0.58 & 0.65 \\

        \bottomrule
    \end{tabular}
    }
    \caption{Classification results with respect to two test sets. P.T.E. stands for the combined model of Profile, Tweets, and Ego. The best scores for each measure and each dataset and the classifiers providing them are bolded.} 
    
    \label{tab:class_results}
\end{table}

\subsection{Models Based on Single Account Data}

\minisection{Profile:} The features that can be created using only the user object, collected from a single tweet or by using the user id: accounts' age, their number of tweets, followers, followings, likes, the length of their name, screen name, description field and a boolean feature indicating whether the account reported a location and a home page.

\minisection{Tweets:} Features we create using the last 200 tweets of the user. We computed the time between our follow request and the users' last tweet to capture activity level (though all users should be active by the time of sampling), the number of tweets with multiple mentions, the average time passed between subsequent tweets, the number of duplicate tweets, the raw number and the percentage of retweets, replies, and quotes, the mean and standard deviation of hashtags and mentions used and likes the tweets got. We also use the content of the tweets by appending the CLS token, employing the ``distilbert-base-multilingual-cased" transformers model due to its efficiency and capability to handle multilingual content as in our case~\cite{sanh2019distilbert}.

\minisection{Ego Network:} The handcrafted features after collecting the followings and followers of a user. These are the ratio of reciprocal relations of the user to their number of followers and followings.

\subsection{Context-Aware (Network) Models}

\minisection{Handcrafted Network Features (H.C.):} Network statistics of the node: (in/out) degree, closeness centrality, (in/out) degree centrality, betweenness, eccentricity, reciprocity, clustering coefficient, PageRank, hub, authority, boolean features indicating whether the node is in a clique or in the GC.

\minisection{Node2Vec:} We train node embeddings following the approach by Grover et al.~\citeyear{grover2016node2vec}. We initially experimented with Japanese and Iranian community to tune the hyper-parameters and found that the parameters prioritizing the local graph structure perform the best for both communities (d = 256, walk length = 40, number of walks = 5, window length = 5, p1 = 1, q = 0.5). We then created the embeddings for the whole network.

\minisection{Graph Neural Networks:} We use profile, tweet metadata and ego network features in combination with the network data using Graph Neural Networks. We tested Graph Convolutional Networks (GCN)~\cite{GCN}, GraphSage~\cite{GraphSage}, and Graph Attention Networks (GAT)~\cite{GAT}. 

\subsection{Experimental Details \& Results}
We employ Sci-Kit Learn's Random Forest implementation with default parameters to do the tabular classifications~\cite{scikit-learn}. The Deep Learning models are trained on a single GPU for 2500 epochs and take 5 minutes at most. 

We found that classifying follow back accounts is challenging due to low recall. While Graph Neural Networks achieve high recall in the Random set, they do not generalize across communities. The best-performing model in the Stratified set is the combined model that relies on single account data that includes the profile, tweet, and ego network. Adding the handcrafted network features to this model slightly improves the results on the Random set. We conclude that Graph Neural Networks with GCN architecture are effective in detecting follow back accounts in large follow back communities (which also have high follow back ratios). Conversely, simpler models generalize better, but they suffer from low recall. 

\section{Conclusions and Implications}
\label{sec:discussion}
In this study, we define, collect a dataset of, characterize, and classify follow back accounts. We conclude by discussing its implications.

\minisection{Implications on Social Media Platforms:} The prevalence of follow-back accounts may undermine the integrity of platforms' user base as it compromises the user metrics such as follower count that are proxies for popularity and credibility. Furthermore, it may poison the recommendation algorithms, promote low-quality content, and disrupt user experience. A network of users with inflated follower counts may also feed the market of popular accounts, which may later be used to undermine civic integrity. Platforms should consider those implications and mitigate the impact of follow back accounts and their communities.

\minisection{Implications on Civic Integrity:} In this study, we revealed 12 follow back communities, many of which exhibit a clear alignment with specific political parties or agendas. In Turkish, Thai, Pakistani, and Indonesian communities, many users have a clear pro-government stance. For instance, the most popular hashtag in the Turkish community (used by 33\% of the users) is \#IStandByMyState (\#DevletiminYanındayım), is actually an astroturfing campaign to counter the opposite voices during the 2021 Boğaziçi protests~\cite{balouglu2021trolls}. The other most popular hashtags such as \#NationalTechnologyMove (\#MilliTeknolojiHamlesi) observe the same pro-government stance while the hashtags \#YouWillBeAccountable (\#HesabınıVereceksiniz) and \#ContinueToSpoilTheGames (\#OyunlarıBozmayaDEVAM) explicitly threaten the opposition. The prevalence in using such polarizing hashtags may suggest that the accounts in this community may be the continuation of the government's efforts to influence the public through coordinated inauthentic accounts. Twitter previously removed and published the data of such inauthentic accounts in 2020, revealing that such accounts showing strong support for the government by using polarizing hashtags in a coordinated way were in fact controlled by the governing party's youth wing~\cite{grossman2020political}. Thus, discovering follow back accounts may help reveal such state-affiliated operational accounts. In the case of US-French, Spanish, and Greek communities, the accounts promote relatively minor parties, i.e., French accounts promote Eric Zemmour who received 7\% in the 2022 presential elections, Spanish accounts promote Vox who received 12\% of the votes in 2023 Spanish elections, and Greek accounts promote PASOK-KINAL who received 11\% of the votes in 2023 Greek elections. Those accounts may be orchestrating astroturfing campaigns to artificially inflate the popularity of certain parties beyond their actual standing. In both cases, the follow back behavior may enhance the popularity and the visibility of the accounts involved, contributing to the increased popularity of the political entities they endorse.

Although follow back accounts are misleading as they inflate their follower counts in a non-organic way that is explicitly prohibited by the platform's terms of service, they may not always have malicious goals. For instance, we found some accounts that grow their circle using follow back although they only share dog photos in the US-FR community. However, such accounts still have the potential to be used for malicious goals later as accounts may alter their profile to be adopted for a malicious campaign later on. Thus, it is still crucial to detect and monitor follow back accounts that purport to be benign users.

\minisection{Potential Misuse of This Work:} Adversaries may use our methodology to find accounts to grow their followers. However, it is important to note that adversaries with the objective of increasing followers are likely already familiar with the follow back strategy. Our work, instead, is more likely to raise awareness among the general public. 

\minisection{Generalizability:} We collect our dataset using seed accounts discovered in the random sample. This made the dataset and the study biased towards the communities of those seed accounts. To mitigate such bias, we include all the communities with a sufficient size, even the ones with very low follow back ratio, report results community-wise, and indicate which communities do not observe the patterns we find. While we acknowledge that the study may not generalize perfectly, our community-oriented approach aims to mitigate the impact of bias to the best extent possible.

\minisection{Limitations:} Our study has limitations mainly due to data restrictions. First, we do not have access to follow the request behavior of individual accounts. This is a limitation that we mitigate by assuming which accounts are follow back accounts and which ones are automated, both justified by our honeypot approach, a method that has proven effective in the past work. Another limitation of this data collection approach is that when our fake account gained followers, they may have built some credibility which may affect some of the accounts' follow back behavior. As it is impractical and unethical to create a new account to follow each account or remove our followers each time we have a follow back (follow churn), we could not mitigate this issue. However, we contend that our accounts remained as blank profiles throughout the experiment and this is likely to have more influence on others' perspectives on our accounts than the few followers they had acquired. 
\section{Ethical Impact}
\label{sec:ethics}

We followed the standard Twitter data analysis practices: we only analyzed public profiles, presented only the aggregate results, and did not manually inspect individuals at any point. Our project is approved by the ethics committee of EPFL with the condition that the subjects should not be deceived and should be informed. Thus, we informed the accounts who followed our blank profiles by sending them a second follow from our "Do Not Follow Back Accounts" and revealing our affiliation. 

In order to comply with Twitter's / X's T.O.S. which prohibits follow churn, we decided not to unfollow the accounts we follow later and keep our accounts as they are. We also decided not to share any data publicly due to privacy concerns and the recent restrictions with accessing and sharing X data. However, researchers can reproduce using our data collection and detection methodology.

\fontsize{9.0pt}{10.0pt} \selectfont
\bibliography{bib}
\subsection{Ethics Checklist}

\begin{enumerate}

\item For most authors...
\begin{enumerate}
    \item  Would answering this research question advance science without violating social contracts, such as violating privacy norms, perpetuating unfair profiling, exacerbating the socio-economic divide, or implying disrespect to societies or cultures? 
    \answerYes{Yes, please see Conclusions and Implications}
  \item Do your main claims in the abstract and introduction accurately reflect the paper's contributions and scope?
    \answerYes{Yes}
   \item Do you clarify how the proposed methodological approach is appropriate for the claims made? 
    \answerYes{Yes, in Data Collection, Characterization, and Classification sections, we state our objective and methods and justify our approach}
   \item Do you clarify what are possible artifacts in the data used, given population-specific distributions?
    \answerYes{Yes, in the Data Collection section when we talk about the snowball sample as well as in the Limitations section}
  \item Did you describe the limitations of your work?
    \answerYes{Yes, please see Conclusions section}
  \item Did you discuss any potential negative societal impacts of your work?
    \answerYes{Yes, the potential negative societal impacts may be deceiving the subjects, which we mitigate as we discuss in the Ethics section, and our method may be misused, which we discuss in the Potential Misuse of }
      \item Did you discuss any potential misuse of your work?
    \answerYes{Yes, please see Ethics section}
    \item Did you describe steps taken to prevent or mitigate potential negative outcomes of the research, such as data and model documentation, data anonymization, responsible release, access control, and the reproducibility of findings?
    \answerYes{Yes, please see Ethics}
  \item Have you read the ethics review guidelines and ensured that your paper conforms to them?
    \answerYes{Yes}
\end{enumerate}

\item Additionally, if your study involves hypotheses testing...
\begin{enumerate}
  \item Did you clearly state the assumptions underlying all theoretical results?
    \answerNA{NA}
  \item Have you provided justifications for all theoretical results?
    \answerNA{NA}
  \item Did you discuss competing hypotheses or theories that might challenge or complement your theoretical results?
    \answerNA{NA}
  \item Have you considered alternative mechanisms or explanations that might account for the same outcomes observed in your study?
    \answerNA{NA}
  \item Did you address potential biases or limitations in your theoretical framework?
    \answerNA{NA}
  \item Have you related your theoretical results to the existing literature in social science?
    \answerNA{NA}
  \item Did you discuss the implications of your theoretical results for policy, practice, or further research in the social science domain?
    \answerNA{NA}
\end{enumerate}

\item Additionally, if you are including theoretical proofs...
\begin{enumerate}
  \item Did you state the full set of assumptions of all theoretical results?
    \answerNA{NA}
	\item Did you include complete proofs of all theoretical results?
    \answerNA{NA}
\end{enumerate}

\item Additionally, if you ran machine learning experiments...
\begin{enumerate}
  \item Did you include the code, data, and instructions needed to reproduce the main experimental results (either in the supplemental material or as a URL)?
    \answerNo{No}
  \item Did you specify all the training details (e.g., data splits, hyperparameters, how they were chosen)?
    \answerYes{Yes, please see Classification Section}
     \item Did you report error bars (e.g., with respect to the random seed after running experiments multiple times)?
    \answerNo{No, our objective is not to create finetuned models but rather to understand what type of model can be helpful to classify follow back accounts}
	\item Did you include the total amount of compute and the type of resources used (e.g., type of GPUs, internal cluster, or cloud provider)?
    \answerYes{Yes, Please see Classification section}
     \item Do you justify how the proposed evaluation is sufficient and appropriate to the claims made? 
    \answerYes{Yes, Please see the Evaluation subsection in the Classification section}
     \item Do you discuss what is ``the cost`` of misclassification and fault (in)tolerance?
    \answerYes{Yes, Please see Classification section}
  
\end{enumerate}

\item Additionally, if you are using existing assets (e.g., code, data, models) or curating/releasing new assets...
\begin{enumerate}
  \item If your work uses existing assets, did you cite the creators?
    \answerYes{Yes}
  \item Did you mention the license of the assets?
    \answerNo{No, as we primarily utilize standard free libraries like sklearn or pytorch, specifying their license would be unnecessary.}
  \item Did you include any new assets in the supplemental material or as a URL?
    \answerNo{No}
  \item Did you discuss whether and how consent was obtained from people whose data you're using/curating?
    \answerYes{Yes, Please see Data Collection and Ethics sections}
  \item Did you discuss whether the data you are using/curating contains personally identifiable information or offensive content?
    \answerYes{Yes, Please see Ethics section}
\item If you are curating or releasing new datasets, did you discuss how you intend to make your datasets FAIR?
\answerNA{NA}
\item If you are curating or releasing new datasets, did you create a Datasheet for the Dataset? 
\answerNA{NA}
\end{enumerate}

\item Additionally, if you used crowdsourcing or conducted research with human subjects...
\begin{enumerate}
  \item Did you include the full text of instructions given to participants and screenshots?
    \answerYes{Yes, please see Data Collection section.}
  \item Did you describe any potential participant risks, with mentions of Institutional Review Board (IRB) approvals?
    \answerYes{Yes, please see Ethics section}
  \item Did you include the estimated hourly wage paid to participants and the total amount spent on participant compensation?
    \answerNA{NA}
   \item Did you discuss how data is stored, shared, and deidentified?
   \answerYes{Yes, please see Ethics section}
\end{enumerate}

\end{enumerate}

\end{document}